\documentclass[pra,superscriptaddress,bibnotes,twocolumn]{revtex4-1}

\usepackage[T1]{fontenc}
\usepackage[utf8]{inputenc}
\usepackage[canadian]{babel}
\usepackage{amsmath,amssymb,bm}
\usepackage{graphicx,xcolor}
\usepackage{titlesec}
\usepackage{microtype}
\usepackage{braket,siunitx}
\usepackage[version=3]{mhchem}
\usepackage[bbgreekl]{mathbbol}
\usepackage{bbm}
\usepackage{braket}
\titleformat*{\paragraph}{\itshape}
\usepackage{float}

\begin{document}
\title{Symmetry conservation with Trotterization and Quantum Phase Estimation}

\author{Edith Leal-S\'{a}nchez}
\affiliation{Department of Chemistry, Dalhousie University, 6243 Alumni Cres, Halifax, NS B3H 4R2, Canada}

\author{Fanny Vain}
\affiliation{Department of Chemistry, Dalhousie University, 6243 Alumni Cres, Halifax, NS B3H 4R2, Canada}

\author{Jong-Kwon Ha}
\affiliation{Department of Chemistry, Dalhousie University, 6243 Alumni Cres, Halifax, NS B3H 4R2, Canada}

\author{Ryan J. MacDonell}
\email{rymac@dal.ca}
\affiliation{Department of Chemistry, Dalhousie University, 6243 Alumni Cres, Halifax, NS B3H 4R2, Canada}
\affiliation{Department of Physics and Atmospheric Science, Dalhousie University, 1493 Lord Dalhousie Dr, Halifax, NS B3H 4R2, Canada}

\begin{abstract}
Quantum algorithms for quantum chemistry and other many-body fermionic systems work by expressing the Hamiltonian in a basis of qubits and fragmenting the Hamiltonian into a sum of products of Pauli operators whose exponentials are easily encoded on a quantum device. Applying the product of exponentials, known as Trotterization, leads to an error associated with the non-commutativity of operators. This error can lead to breaking the symmetries of the Hamiltonian because the fragments are not symmetry conserving in general. Nonetheless, many algorithms for time evolution rely on Trotterization, including time evolution and quantum phase estimation. We show that we can express the Hamiltonian in terms of Hermitian excitation operators which map to sums of commuting Pauli strings for any encoding and conserve symmetries corresponding to Abelian groups of symmetry operators. Symmetries corresponding to non-Abelian groups, on the other hand, are not fully conserved by Trotterized Hermitian excitation operators, so we developed ``operator kirigami'' to cut the sum of non-commuting operators by orthogonal projection and to fold terms together using unitary rotations. We tested pools of operators for small molecules and basis sets, and found that electron number and spin symmetry conserving pools led to greater errors that decreased for larger molecules and were negated with second-order Trotterization. Our work shows the potential for testing quantum computing algorithms on classical computers by adapting tools used in electronic structure theory with conserved symmetries.
\end{abstract}
\maketitle

The primary challenge for the simulation of quantum systems and materials is the exponential growth of the Hilbert space with increasing degrees of freedom, which quickly reaches the limits of classical computers. Several approximations, such as mean-field theories, density functional theories, and tensor-network methods, have enabled the simulation of a wide range of systems in the past century~\cite{murg2015,schollwock2005,schirmer2025,pederson2022}. Quantum computing offers a new paradigm for physical simulation: quantum computers span Hilbert spaces that scale exponentially with the number of qubits, but relative to classical computing algorithms, information about the state is more difficult to obtain~\cite{Nielsen,garcia2021}. Recent years have seen an ever increasing interest in developing practical quantum computing algorithms, with physical simulation being the most promising early application~\cite{noh2016,daley2022,childs2018,lee23,ollitrault2023}.

Several quantum computing algorithms have emerged as a basis for physical simulation. For near-term applications, quantum computers are limited by their short decoherence times, requiring low-depth algorithms with classical feedback. The variational quantum eigensolver (VQE)~\cite{peruzzo2014} and its many variations~\cite{cerezo2021,grimsley2019,tilly2022} provides such an algorithm. However, VQE simulations for realistic systems are severely limited by the number of measurements they require~\cite{gonthier2022,huggins2021,verteletskyi2020}. For long-term applications with fault-tolerant quantum computers, quantum phase estimation (QPE) offers a straightforward approach to find eigenstates and properties of quantum systems~\cite{aspuru-guzik2005,mcardle2020,babbush2018}. Both VQE and QPE rely on writing the Hamiltonian as a sum of easily simulated fragments. The Trotter approximation~\cite{trotter1959} is used to simulate the dynamics of the system in order to find a time-dependent phase in QPE, and Trotter products are a common parametrization for variational quantum algorithms. More generally, the Trotter approximation can be used to simulate spectra and other time-dependent observables of quantum systems~\cite{abrams1997,lloyd1996}.

Quantum computing encodings of fermions work by mapping the full Fock space of a set of spin orbitals onto the Hilbert space of an equal~\cite{jordan1928,bravyi2002,seeley2012} or greater~\cite{bravyi2002sf,setia2018} number of qubits. In contrast, electronic structure algorithms for classical computers take advantage of universal Hamiltonian symmetries to perform calculations in a subspace of the multi-electron Fock space~\cite{helgaker,szabo}. The electronic Hamiltonian commutes with the electron number operator, and in the absence of spin-orbit coupling, the spin-projection operator. The total spin operator commutes with the electronic Hamiltonian if there is also no external magnetic field present~\cite{helgaker,szabo}. Electronic structure algorithms on classical computers significantly reduce the multi-electron basis to those with pre-defined quantum numbers, enabling large-scale calculations.
Recent work has focused on the symmetries involved in quantum computing algorithms, including particle and spin symmetries~\cite{jain2026, magoulas2026,sung2026} and geometric symmetries~\cite{picozzi2023}. Additional work has shown how qubit counts can be reduced using symmetries~\cite{harrison2022}, albeit at a logarithmic scale relative to the reduction in the cost of classical computation.

\begin{figure*}
    \centering
    \includegraphics{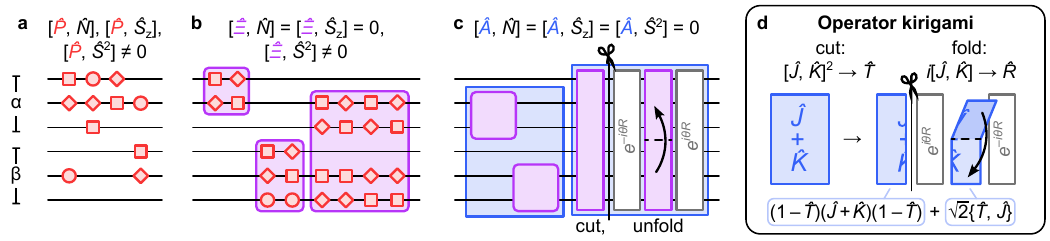}
    \caption{Summary of Hamiltonian fragmentation schemes and their conservation of symmetries with Trotterization. \textbf{a}, The Hamiltonian is decomposed into a sum of Pauli strings ($\hat{P}$) with Pauli operators (red shapes) acting on qubits (black lines). The qubits represent spin orbitals with spins $\alpha$ and $\beta$. For random orders of Pauli strings, symmetries such as electron number ($\hat{N}$) and spin ($\hat{S}_z$, $\hat{S}^2$) are not conserved. \textbf{b}, Hermitian excitation operators ($\hat{\Xi}$) generate groups of Pauli strings that mutually commute, thus conserving electron number, spin projection, and other symmetries given by Abelian groups. \textbf{c}, Operators which conserve symmetries corresponding to non-Abelian groups (such as total spin) are composed of sums of Hermitian excitation operators ($\hat{A}$), some which haver terms that do not commute. The non-commuting terms can be resolved through operator kirigami, introduced in this paper. \textbf{d}, Operator kirigami involves a \emph{cut} by an orthogonal projection operator $\hat{T}$ and a \emph{fold} by a unitary rotation operator $\hat{R}$ with an angle $\theta$. The projection and rotation operators are both derived from the commutators of Hermitian excitation operators ($\hat{J}$, $\hat{K}$) in the fragment.}
    \label{fig:summary}
\end{figure*}

In this paper, we show that all fermion-to-qubit mappings (Fig.~\ref{fig:summary}a) have a common set of Hermitian excitation operators that conserve symmetries corresponding to an Abelian group, such as electron number and spin projection symmetries, with Trotterization (Fig.~\ref{fig:summary}b). We find that fragmentation of the Hamiltonian alone is not enough to conserve symmetries corresponding to non-Abelian groups, such as total spin symmetry. Instead, we develop a technique that we call ``operator kirigami'': the properties of Hermitian excitation operators allow us to \emph{cut} symmetry-conserving fragments through orthogonal projection, and to \emph{fold} non-commuting terms into a single operator with unitary rotations. Both operations are derived from the commutator of Hermitian excitation operators in the fragment (Fig.~\ref{fig:summary}d). Implementation of the symmetry conserving operator amounts to cutting and unfolding the folded term (Fig.~\ref{fig:summary}c). We conclude by demonstrating the consequences of spin symmetry conservation for time-evolution and QPE with a set of molecules and basis sets.

Quantum dynamics simulations on quantum computers, including QPE, employ the Trotter approximation for time evolution with a Hamiltonian by expressing it as a sum of fragments, $\hat{H} = \sum_j \hat{H}_j$, then approximating the exponential of the Hamiltonian as a product of fragment exponentials,
\begin{equation} \label{eq:trotter}
    \exp(\hat{H}\tau) \approx \exp(\hat{H}_\mathrm{eff}\tau) = \prod_j \exp(\hat{H}_j \tau),
\end{equation}
where $\hat{H}_\mathrm{eff}$ is an effective Hamiltonian, and $\tau = -i\Delta t$ is an imaginary timestep. We use atomic units ($\hbar = 4\pi\varepsilon_0 = m_e$) here and throughout the manuscript. The error introduced by Trotterization is given by the Baker-Campbell-Hausdorff (BCH) equation~\cite{mehendale2025},
\begin{multline} \label{eq:bch}
    \hat{H}_\mathrm{eff} = \hat{H} + \frac{\tau}{2} \sum_{j > k} [\hat{H}_j, \hat{H}_k] \\
+ \frac{\tau^2}{6} \sum_{j \ge k \ge l} \biggl((1 - \frac{\delta_{jk}}{2}) [\hat{H}_j, [\hat{H}_k, \hat{H}_l]] \\ +
(1 - \frac{\delta_{kl}}{2}) [\hat{H}_l, [\hat{H}_k, \hat{H}_j]]\biggr) + \mathcal{O}(\tau^3).
\end{multline}
The Trotter error arises only for fragments that do not commute, but the error does not readily truncate at low orders~\cite{childs2021}. Much of the focus on error reduction for these algorithms has thus centered on grouping commuting fragments to reduce errors~\cite{mehendale2025,verteletskyi2020}, and randomizing orders of fragments or groups such that errors approximately cancel~\cite{campbell2019}.

To obtain Hamiltonian fragments for a set of fermions, we must adopt a fermion-to-qubit encoding. Such an encoding must obey the anti-commutation rules $\{\hat{a}_p,\hat{a}_q\} = \delta_{pq}$ and $\{\hat{a}_p,\hat{a}_q\} = \{\hat{a}_p^\dag,\hat{a}_q^\dag\} = 0$ where $\hat{a}_p^\dag$ and $\hat{a}_p$ are fermion creation and annihilation operators for spin-orbital $p$, respectively. Pauli operators $\hat{X}$, $\hat{Y}$ and $\hat{Z}$ form a basis for qubit operations, and their tensor products (Pauli strings) form a basis of Hermitian, unitary operators in the qubit state space. Majorana operators, given by $\hat{\gamma}_p^+ = \hat{a}_p^\dagger + \hat{a}_p$ and $\hat{\gamma}_p^- = i(\hat{a}_p^\dagger - \hat{a}_p)$, thus map naturally to Pauli strings~\cite{loaiza2025}. For example, encodings with one qubit per spin orbital can be written as
\begin{align}
    \hat{\gamma}_p^+ &= \hat{a}_p^\dagger + \hat{a}_p \mapsto \hat{X}_{\mu(p)} \hat{X}_{p} \hat{Z}_{\pi(p)} \label{eq:majorana0}, \\
    \hat{\gamma}_p^- &= i(\hat{a}_p^\dagger - \hat{a}_p) \mapsto \hat{X}_{\mu(p)} \hat{Y}_p \hat{Z}_{\phi(p)} \label{eq:majorana1},
\end{align}
where $\mu(p)$ is a collective index for spin orbital indices greater than $p$, whereas $\pi(p)$ and $\phi(p)$ are collective indices for indices less than $p$. The non-local part of the Majorana operators encodes the parity, defined by the sum (mod 2) of occupations of orbitals with indices less than $p$~\cite{loaiza2025}. Each fermion-to-qubit encoding adopts different sets of indices for $\mu(p)$, $\pi(p)$ and $\phi(p)$.
If we write two Hermitian excitation operators,
\begin{align}
    \hat{\Xi}_q^p &= \hat{\Xi}_p^q = \hat{a}_p^\dag \hat{a}_q + \hat{a}_q^\dag \hat{a}_p = \frac{1}{2}\left(\hat{\gamma}_p^+ \hat{\gamma}_q^+ + \hat{\gamma}_p^- \hat{\gamma}_q^-\right), \\
    \hat{\Sigma}_q^p &= -\hat{\Sigma}_p^q = i(\hat{a}_p^\dag \hat{a}_q - \hat{a}_q^\dag \hat{a}_p) = \frac{1}{2}\left(\hat{\gamma}_p^- \hat{\gamma}_q^+ - \hat{\gamma}_p^+ \hat{\gamma}_q^-\right),
\end{align}
then the operators each map to a pair of commuting Pauli strings with real coefficients for any encoding with one Pauli string per Majorana operator. Although in principle other encodings could be devised, we assume one Pauli string per Majorana operator in the remainder of this paper since it is true of all common fermion-to-qubit encodings.

To include electron spin, we add an index $\sigma$ to distinguish spin orbitals with the same spatial orbital. This separates the qubits into two registers with $\sigma = \alpha$ for spin $+\tfrac{1}{2}$ and $\sigma = \beta$ for spin $-\tfrac{1}{2}$. Without spin-orbit coupling or an external magnetic field, the electronic Hamiltonian can be written as
\begin{equation} \label{eq:ham}
    \hat{H} = \frac{1}{2}\sum_{pq} \tilde{k}_q^p \sum_\sigma \hat{\Xi}_{q\sigma}^{p\sigma} + \frac{1}{16}\sum_{pqrs} \tilde{v}_{rs}^{pq} \sum_{\sigma\sigma'}  \{\hat{\Xi}_{r\sigma}^{p\sigma},\hat{\Xi}_{s\sigma'}^{q\sigma'}\},
\end{equation}
where $\tilde{k}_q^p$  and $\tilde{v}_{rs}^{pq}$ are coefficients given by integrals over spatial orbitals (provided in Appendix~\ref{app:integham}). The Hamiltonian has several symmetries corresponding to groups of operators that commute with the Hamiltonian. For example, the electron number is conserved because the number operator $\hat{N} = \sum_p \sum_\sigma \hat{\Xi}_{p\sigma}^{p\sigma}$ commutes with each term. The spin projection operator $\hat{S}_z = \frac{1}{4}\sum_p (\hat{\Xi}_{p\alpha}^{p\alpha} - \hat{\Xi}_{p\beta}^{p\beta})$ also commutes with each excitation operator $\hat{\Xi}_{p\sigma}^{q\sigma}$. Each of the two-electron terms $\{\hat{\Xi}_{r\sigma}^{p\sigma},\hat{\Xi}_{s\sigma'}^{q\sigma'}\}$ also maps to a sum of commuting Pauli strings by construction, and thus each term in Eq.~\ref{eq:ham} conserves electron number and spin projection. More generally, all symmetries corresponding to Abelian groups (i.e. groups where all operators commute) are conserved by non-vanishing terms in the form of Eq.~\ref{eq:ham} if the orbitals are symmetry adapted such that each orbital belongs to an irreducible representation (see Appendix~\ref{app:integham}).

Symmetries corresponding to non-Abelian groups are not conserved by individual terms in Eq.~\ref{eq:ham}. For the remainder of this paper we focus on the non-Abelian group with $\hat{N}$, $\hat{S}_z$ and the remaining spin operators $\hat{S}_x$ and $\hat{S}_y$; however, the techniques we develop are broadly applicable. The operators $\hat{S}_x = \frac{1}{2} \sum_p \hat{\Xi}_{p\beta}^{p\alpha}$ and $\hat{S}_y = -\frac{1}{2} \sum_p \hat{\Sigma}_{p\beta}^{p\alpha}$ each involve excitations between degenerate spin orbitals. Evaluating their commutators reveals $[\hat{S}_x,\hat{\Xi}_{q\alpha}^{p\alpha}] = -[\hat{S}_x,\hat{\Xi}_{q\beta}^{p\beta}]$ and $[\hat{S}_y,\hat{\Xi}_{q\alpha}^{p\alpha}] = -[\hat{S}_y,\hat{\Xi}_{q\beta}^{p\beta}]$,
meaning the operator $\hat{A}_q^p = \sum_\sigma \hat{\Xi}_{q\sigma}^{p\sigma}$ conserves total spin. For all two-electron operators with non-intersecting sets of indices ($\{p\sigma,r\sigma\} \cap \{q\sigma',s\sigma'\} = \varnothing$), the Pauli strings generated from $\hat{\Xi}_{r\sigma}^{p\sigma}$ commute with the Pauli strings from $\hat{\Xi}_{s\sigma'}^{q\sigma'}$ due to the fermion anti-commutation rules, and total spin is thus conserved for those terms.

Only two types of terms arise from two-electron operators expressed as products of sums of Hermitian excitation operators that do not commute, which we label $\hat{B}_q^p$ and $\hat{C}_q^{pr}$ (for all unique terms, see Appendix~\ref{app:integham}). The first has two unique indices, and is given by
\begin{multline} \label{eq:beq}
    \hat{B}_q^p = \frac{1}{2}\{\hat{\Xi}_{q\alpha}^{p\alpha} + \hat{\Xi}_{q\beta}^{p\beta}, \hat{\Xi}_{q\alpha}^{q\alpha} + \hat{\Xi}_{q\beta}^{q\beta}\} \\
    = \left[\hat{\Xi}_{q\alpha}^{p\alpha} + \hat{\Xi}_{q\beta}^{p\beta}\right] - \left[\hat{\Xi}_{q\alpha}^{p\alpha}(1-\hat{\Xi}_{q\beta}^{q\beta}) + \hat{\Xi}_{q\beta}^{p\beta}(1-\hat{\Xi}_{q\alpha}^{q\alpha})\right],
\end{multline}
which makes use of the fact that $\hat{\Xi}_{q\sigma}^{p\sigma}(1 - \hat{\Xi}_{q\sigma}^{q\sigma}) = i\hat{\Sigma}_{q\sigma}^{p\sigma}$ is anti-Hermitian (see Appendix~\ref{app:prop}). The operator $1 - \hat{\Xi}_{q\sigma}^{q\sigma}$ is composed of a single Pauli string for all fermion-to-qubit encodings. As a result, each operator in the square brackets is spin conserving and maps to commuting Pauli strings, and can thus be exponentiated individually. In contrast, the non-commuting product with three unique indices is not resolvable into spin-conserving groups of Pauli strings. It can be written as three fragments,
\begin{equation}
    \hat{C}_q^{pr} = \frac{1}{2}\{\hat{\Xi}_{q\alpha}^{p\alpha} + \hat{\Xi}_{q\beta}^{p\beta}, \hat{\Xi}_{r\alpha}^{q\alpha} + \hat{\Xi}_{r\beta}^{q\beta}\} = \hat{G}_0 + \hat{G}_+ + \hat{G}_-,
\end{equation}
where $\hat{G}_+ = \hat{\Xi}_{q\alpha}^{p\alpha}\hat{\Xi}_{r\beta}^{q\beta}$, $\hat{G}_- = \hat{\Xi}_{q\beta}^{p\beta}\hat{\Xi}_{r\alpha}^{q\alpha}$ and $\hat{G}_0 = (\hat{G}_{0\alpha} + \hat{G}_{0\beta}) / 2$ with sub-fragments $\hat{G}_{0\sigma} = \hat{\Xi}_{r\sigma}^{p\sigma}(1 - \hat{\Xi}_{q\sigma}^{q\sigma})$.
Each fragment is encoded as four commuting Pauli strings, but the fragments do not commute with each other. The operators $\hat{G}_\pm$ cannot be further sub-divided because all four Pauli strings are necessary for electron number conservation.
Furthermore, we ruled out the possibility of conserving total spin by adding and subtracting Pauli strings to the groups by testing all sets of Hermitian excitation operators and their products in a basis of six spin orbitals.

In order to implement non-Abelian symmetry conserving operators, we developed a technique to \emph{cut} a fragment into commuting and non-commuting components, and then \emph{fold} the non-commuting component into a single Hermitian excitation operator, which we call ``operator kirigami'' based on the paper-cutting art form (details given in Appendix~\ref{app:kirigami}). The technique uses the fact that all Hermitian excitation operators $\hat{J}$ are tripotent ($\hat{J} = \hat{J}^3$), and they have a block-monomial form in a basis of Slater determinants with all $2^{2N_\mathrm{orb}}$ occupations for $N_\mathrm{orb}$ spatial orbitals. All anticommutators of Hermitian excitation operators ($\{\hat{\Xi}_{r\sigma}^{p\sigma},\hat{\Xi}_{s\sigma'}^{q\sigma'}\}$) and all commutators multiplied by the imaginary unit $i$ ($i[\hat{\Xi}_{r\sigma}^{p\sigma},\hat{\Xi}_{s\sigma'}^{q\sigma'}]$) have the same form, giving a generalization for $n$-electron excitations. The tripotent nature of the operators leads to a simple expression for their exponential,
\begin{equation}
    \exp(i\theta\hat{J}) = 1 - \hat{J}^2 + \cos(\theta)\hat{J}^2 + i\sin(\theta)\hat{J},
\end{equation}
which implies that the vector space of the projection $1 - \hat{J}^2$ remains unchanged, whereas the remaining space is transformed by $\hat{J}$.

\begin{table*}
    \centering
    \caption{Characteristics of each of the simulated molecules. The number of spatial orbitals, number of electrons, and timestep are given by $N_\mathrm{orb}$, $N$ and $\Delta t$, respectively. The number of qubits is equal to $2N_\mathrm{orb}$ in all cases. H$_2$O uses the frozen core approximation. $N_\mathrm{oper}$ is the number of operators in each operator pool. ``Dim.'' is the size of the minimum state space given the symmetries of each operator pool.}
    \begin{tabular*}{\textwidth}{@{\extracolsep{\fill}} |c|c|c|c|c|c|c|c|c|c|c|c| @{}}
    \hline 
    \multicolumn{6}{|c|}{}  & \multicolumn{2}{c|}{Qubit} & \multicolumn{2}{c|}{Slater} & \multicolumn{2}{c|}{CSF}  \\
    \hline
    System & Basis & $N_\mathrm{orb}$ & $N$ & $\Delta t$ / a.u. & $E_\mathrm{HF}-E_\mathrm{CI}$ / $E_h$  &  $N_\mathrm{oper}$ & Dim. &  $N_\mathrm{oper}$ & Dim. &  $N_\mathrm{oper}$ & Dim. \\
    \hline
    H$_2$ & STO-3G  & 2 & 2 & 0.001 & 0.0205 &  15 & 16 & 9 & 4 & 9 & 3 \\
          & 6-31G   & 4 & 2 & 0.08  & 0.0249 &  185  & 256 & 89 & 16  & 77 & 10 \\
          & cc-pVDZ & 6 & 2 & 0.05  & 0.0347 &  919 & 4096 & 372 & 36 & 312 &  21 \\ \hline
   LiH    & STO-3G  & 4 & 4 & 0.05  & 0.0201 &  361 & 256 & 89 & 36 & 77 &  20 \\
          & 6-31G   & 7 & 4 & 0.005 & 0.0189 &  3382 & 16384 & 644 & 441 & 539 &  196 \\ \hline
   H$_2$O & STO-3G  & 6 & 8 & 0.004 & 0.0517 & 1086 & 4096 & 372 & 225 & 312 & 105 \\ \hline
    \end{tabular*}
    \label{tab:results}
\end{table*}

For two non-commuting Hermitian excitation operators $\hat{J}$ and $\hat{K}$, rotation of either operator by $\hat{L} = i[\hat{J},\hat{K}]$ with an angle $\theta$ leads to a linear combination $\hat{J}$ and $\hat{K}$ in the space given by $\hat{L}^2$. For a linear combination of $\hat{J}$ and $\hat{K}$ with real coefficients $c_J$ and $c_K$, we can write
\begin{multline}
    \exp(i\theta\hat{L}) (c_J \hat{J} + c_K \hat{K}) \exp(-i\theta\hat{L}) \\
    = (1-\hat{L}^2)(c_J\hat{J} + c_K\hat{K})(1-\hat{L}^2) + \sqrt{c_J^2 + c_K^2} \{\hat{L}^2,\hat{J}\},
\end{multline}
where $\theta = \arctan(-c_K/c_J)$. $1 - \hat{L}^2$ defines the commuting space of $\hat{J}$ and $\hat{K}$, since $(1 - \hat{L}^2) \hat{L} = 0$. The exponential cuts (projects) the linear combination into commuting and non-commuting spaces, and folds (rotates) the non-commuting component into a single operator. The folded operator can be unfolded during implementation to achieve the linear combination of non-commuting terms without Trotter error.

Applying operator kirigami to the $\hat{C}_q^{pr}$ operator, we first evaluated $i[\hat{G}_+,\hat{G}_-]$ to simplify $\hat{G}_+ + \hat{G}_-$ into commuting terms. Applying the same operation to $\hat{G}_0$ yielded two excitation operators corresponding to $\hat{G}_{0\alpha}$ and $\hat{G}_{0\beta}$. However, due to the null space of the fragment operators, we found that we could simplify the transformation by isolating a single term from the commutator. This simplifies the folding operator to $\hat{R}_+ = \hat{\Sigma}_{r\alpha}^{p\alpha} \hat{\Xi}_{r\beta}^{p\beta}(1 - \hat{\Xi}_{q\beta}^{q\beta})$ such that
\begin{align}
    \exp&\left(-i\frac{\pi}{4}\hat{R}_+\right) \hat{C}_q^{pr} \exp\left(i\frac{\pi}{4}\hat{R}_+\right) \nonumber\\
    &= (1 - \hat{R}_+^2) \hat{C}_q^{pr} (1 - \hat{R}_+^2) + \sqrt{2}\{\hat{R}_+^2,\hat{G}_+\} + \hat{\Gamma}_\beta, \\
    \hat{\Gamma}_\beta &= \frac{1}{4}\hat{G}_{0\beta} \left(\hat{\Xi}_{p\alpha}^{p\alpha}(2 - \hat{\Xi}_{q\alpha}^{q\alpha} - \hat{\Xi}_{r\alpha}^{r\alpha}) + \hat{\Xi}_{q\alpha}^{q\alpha}\hat{\Xi}_{r\alpha}^{r\alpha}\right)
\end{align}
where $\Gamma_\beta$ is in the form of a Hermitian excitation operator. By evaluating $i[\{\hat{R}_+^2,\hat{G}_+\},\hat{\Gamma}_\beta]$, we were again able to isolate a single term, giving $\hat{R}_0 = \hat{\Sigma}_{q\alpha}^{p\alpha} (1 - \hat{\Xi}_{r\alpha}^{r\alpha})\hat{\Xi}_{q\beta}^{p\beta}$. Furthermore, we found that for both operations the projection could be simplified to $\hat{T} = (1 + \prod_\sigma (1 - \hat{\Xi}_{p\sigma}^{p\sigma})(1 - \hat{\Xi}_{r\sigma}^{r\sigma}))/2$, which commutes with $\hat{G}_0$, $\hat{G}_+$ and $\hat{G}_-$. This yields the equation
\begin{equation} \label{eq:decomp}
    \hat{C}_q^{pr} = \hat{C}_q^{pr}(1 - \hat{T}) + \hat{U}_+^\dagger \hat{U}_0^\dagger (\sqrt{3}\hat{\Gamma}_\beta) \hat{U}_0 \hat{U}_+,
\end{equation}
where $\hat{U}_+ = \exp(-i\frac{\pi}{4}\hat{R}_+)$, $\hat{U}_0 = \exp(i\theta_m\hat{R}_0)$, and $\theta_m = \arctan(\sqrt{2})$ is the magic angle (Further details given in Appendix~\ref{app:kirigami}).

The decomposition of $\hat{C}_q^{pr}$ is not only spin conserving with Trotterization, but it also exactly implements $\exp(\hat{C}_q^{pr}\tau)$ with no Trotter error. The commutators of the fragments have the relations $[\hat{G}_+,\hat{G}_-]\hat{T} = [\hat{G}_+,\hat{G}_-]$ and $[\hat{G}_0,\hat{G}_\pm]\hat{T} = [\hat{G}_0,\hat{G}_\pm]$, which results in the Trotter errors cancelling to every order in the BCH expansion for exponentials of $\hat{G}_0 (1 - \hat{T})$ and $\hat{G}_\pm (1 - \hat{T})$. In the space defined by $\hat{T}$, each term is an exponential of a Hermitian excitation operator composed of commuting Pauli strings. By definition, the projections $\hat{T}$ and $1 - \hat{T}$ commute, and thus all Trotter error is negated. Using operator kirigami, we also find that the Trotter error for $\hat{B}_q^p$ can be negated. Operator kirigami is broadly applicable to reduce or remove Trotter error, but the required compilation has a cost that grows exponentially with the number of Hermitian excitation operators. Symmetries result in static relations between operators, making Eq.~\ref{eq:decomp} applicable to any electronic Hamiltonian that conserves electron number and total spin. Further work is required to find simplifications for higher symmetries, such as spatial triple degeneracies and combined spin and spatial symmetry.

Symmetry conservation has an exponentially greater impact on resource requirements for classical computing algorithms relative to quantum computing~\cite{harrison2022,picozzi2023}. For spin symmetries, a system with $N$ electrons in $N_\mathrm{orb}$ spatial orbitals and a spin projection $S_z$ has $N_\mathrm{SD} = \prod_\sigma \binom{N_\mathrm{orb}}{N_\sigma}$ Slater determinants, where $N_\alpha = N/2 + S_z$ and $N_\beta = N/2 - S_z$. Using a total spin $S$ yields a set of configuration state functions (CSFs) whose total count is
\begin{equation}
    N_\mathrm{CSF} = \frac{2S + 1}{N_\mathrm{orb}+1} \binom{N_{\mathrm{orb}}+1}{N/2-S} \binom{N_\mathrm{orb}+1}{N/2+S+1}.
\end{equation}
Assuming the number of orbitals scales with the number of electrons, $N_\mathrm{orb} = c N$ with a  constant $c > 1$, the ratio $2^{2N_\mathrm{orb}} / N_\mathrm{SD}$ is exponential, whereas $N_\mathrm{SD} / N_\mathrm{CSF}$ is linear in the large $N$ limit. Nonetheless, the linear reduction in cost can make the difference between tractable and intractable calculations. Aside from physical correctness, we see the ability to conserve symmetries as a means to translate quantum computing algorithms to a CSF basis in order to test quantum computing algorithms at the same scale as electronic structure on classical computers.

\begin{figure*}
   \centering
   \includegraphics{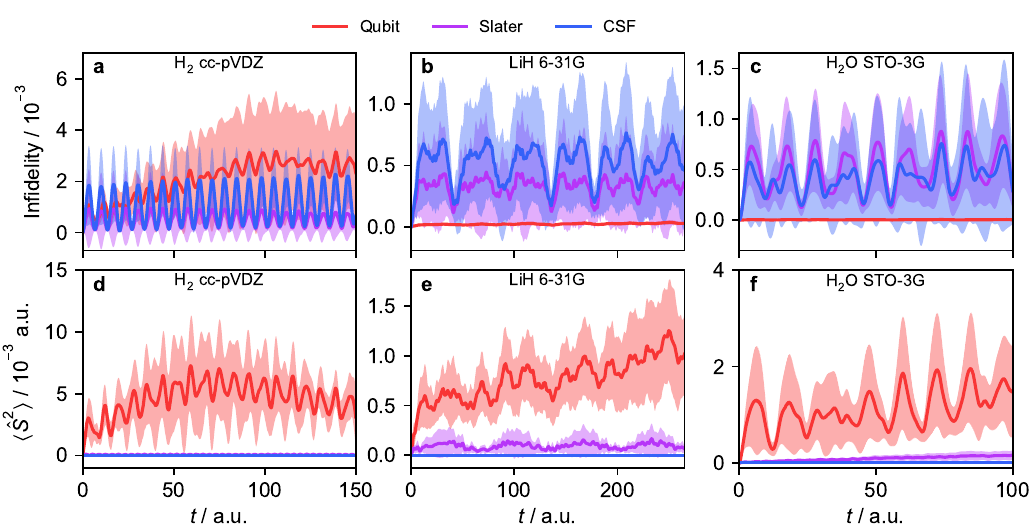}
   \caption{Time evolution of Trotter errors for molecular simulations with the different operator pools. \textbf{a}--\textbf{c}, The average infidelities, $1 - |\braket{\Psi_\mathrm{Trot}|\Psi_\mathrm{exact}}|$, and their standard deviations for 10 random orders of operators. \textbf{d}--\textbf{f}, The average spin squared expectation values, $\braket{\hat{S}^2}$, and their standard deviations for 10 random orders of operators. Columns represent different molecules and basis sets: H$_2$ with a cc-pVDZ basis (\textbf{a}, \textbf{d}), LiH with a 6-31G basis (\textbf{b}, \textbf{e}), and H$_2$O with a STO-3G basis (\textbf{c}, \textbf{f}). All axes share the same units, but note the differences in scales.
   }
    \label{fig:time_ev}
\end{figure*}

To demonstrate the impact of spin symmetry conservation, we performed electronic dynamics simulations for several small molecules and basis sets. For all simulations, we optimized the Hartree-Fock (HF) ground-state minimum geometry, then used the HF orbitals and occupations as a starting point, $\ket{\Psi(0)}$, for dynamics with the full configuration interaction (CI) Hamiltonian and for calculations of ground state energies with Trotter error. Electronic integrals, electronic structure calculations, and geometry optimizations were all performed with PySCF~\cite{pyscf,geometric}. We defined three pools of operators: `Qubit', consisting of all unique Pauli strings in the Jordan-Wigner encoded Hamiltonian; `Slater', consisting of all unique Hermitian excitation operators $\hat{\Xi}_{q\sigma}^{p\sigma}$ and $\{\hat{\Xi}_{r\sigma}^{p\sigma},\hat{\Xi}_{s\sigma'}^{q\sigma'}\}$; and `CSF', consisting of all $\hat{A}_q^p = \sum_\sigma \hat{\Xi}_{q\sigma}^{p\sigma}$ and $\{\hat{A}_r^p,\hat{A}_s^q\}$ with the exception of $\hat{B}_q^p$ (Eq.~\ref{eq:beq}), which was implemented as two spin conserving terms $\hat{A}_q^p$ and $\hat{B}_q^p - \hat{A}_q^p$.

Table~\ref{tab:results} shows the characteristics for the simulations of the three different molecules (H$_2$, LiH, and H$_2$O) and basis sets (STO-3G, 6-31G, cc-pVDZ) used in our simulations. For LiH and H$_2$, we exploited the linear symmetry to only include completely symmetric s and p$_z$ basis functions without loss of accuracy. For H$_2$O, we used the frozen-core approximation to remove the configurations generated by the excitations of the electrons in the oxygen 1s orbital. For dynamics simulations, we selected a timestep $\Delta t$ which was sufficient to accurately measure the correlation energy $E_\mathrm{HF} - E_\mathrm{CI}$, where $E_\mathrm{HF} = \braket{\Psi(0)|\hat{H}|\Psi(0)}$ and $E_\mathrm{CI}\ket{\Psi_0} = \hat{H}\ket{\Psi_0}$ for ground state $\ket{\Psi_0}$. In the final columns of Table~\ref{tab:results} we show the number of operators ($N_\mathrm{oper}$) and the dimension of the full Hilbert space, the space of Slater determinants with fixed $N$ and $S_z$, and the CSF space with fixed $N$ and $S$. All examples had $S_z = S = 0$. Even for the small systems considered, symmetry conservation shows a drastic reduction in the number of operators and the dimension. The most extreme example is LiH with a 6-31G basis, where the number of operators reduces by a factor of 6 and the dimension reduces by a factor of 84.

\begin{figure*}
    \centering
    \includegraphics{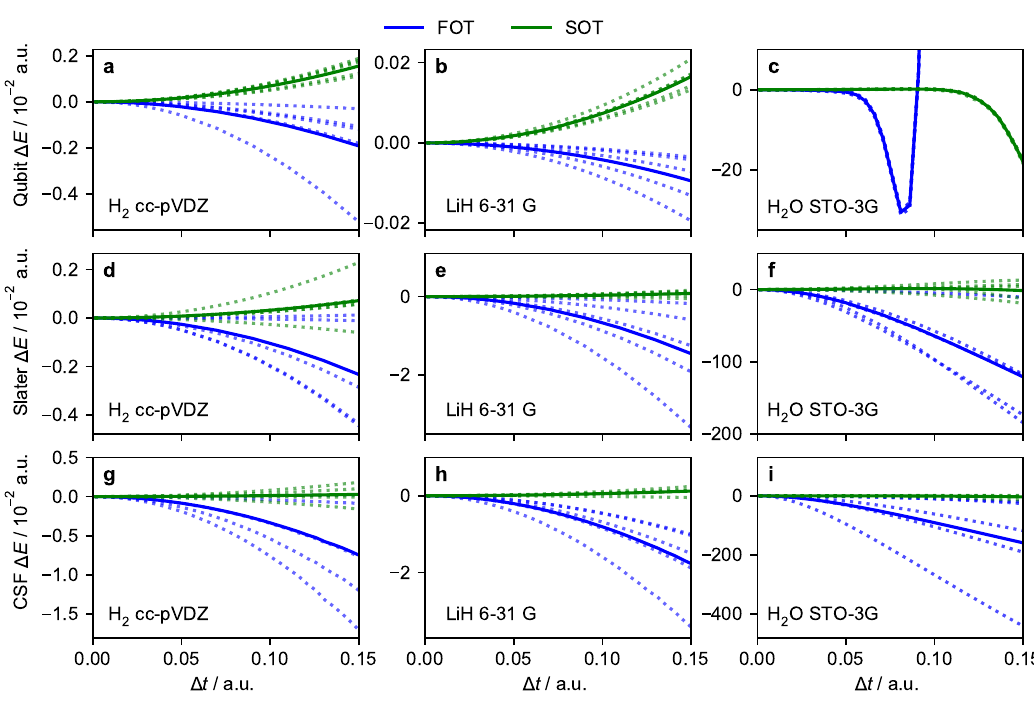}
    \caption{Energy differences between the full CI electronic ground state energy and the Trotter energy for different molecules and operator pools. Five random orders of operators are show with dashed lines, and their average is show as a solid line for first-order Trotter (FOT) and second-order Trotter (SOT). Rows represent different operator pools: Qubit (\textbf{a}--\textbf{c}), Slater (\textbf{d}--\textbf{f}), and CSF (\textbf{g}--\textbf{i}). Columns represent different molecules and basis sets: H$_2$ with a cc-pVDZ basis (\textbf{a}, \textbf{d}, \textbf{g}), LiH with a 6-31G basis (\textbf{b}, \textbf{e}, \textbf{h}), and H$_2$O with a STO-3G basis (\textbf{c}, \textbf{f}, \textbf{i}). All axes share the same units, but note the differences in energy scales.
    }
    \label{fig:log}
\end{figure*}

To understand the error accumulation with Trotterization, we performed first-order Trotter dynamics simulations with each operator pool. We employed a number of timesteps equal to $5 / \Delta t(E_{HF}-E_{CI})$ to simulate a realistic timescale for ground state energy estimation. Each simulation was repeated with 10 random orders of operators, with the order being fixed for all timesteps in a simulation. Fig.~\ref{fig:time_ev} shows the infidelities ($1-|\braket{\Psi_\mathrm{Trot}|\Psi_\mathrm{exact}}|$, where $\ket{\Psi_\mathrm{exact}} = e^{\hat{H}\tau}\ket{\Psi(0)}$ and $\ket{\Psi_\mathrm{Trot}} = e^{\hat{H}_\mathrm{eff}\tau}\ket{\Psi(0)}$) and the $\hat{S}^2 = \hat{S}_x^2 + \hat{S}_y^2 + \hat{S}_z^2$ expectation values and their standard deviations as a function of time. Results using other basis sets as well as $\hat{N}$ and $\hat{S}_z$ expectation values are provided in Appendix~\ref{app:ap3}. As expected, the Qubit pool shows a faster growth in $\langle\hat{S}^2\rangle$ than the Slater pool, since the latter is exclusively from the terms in $\hat{B}_q^p$ and $\hat{C}_q^{pr}$. However, despite conserving physical symmetries, the Slater and CSF pools have a faster increase in the infidelity for LiH and H$_2$O. We attribute the greater error to the restriction in the number of operators. Each symmetry conserving operator can be written as a sum of Pauli strings, and thus the commutators between operators is the sum of many Pauli string commutators. Random orders of Pauli strings have a greater sign variation, leading more possibilities of Trotter error cancellation.

To investigate the effects of Trotter error on QPE energies, we evaluated the energy error given by $\Delta E = E_\mathrm{eff} - E_\mathrm{CI}$, where $E_\mathrm{eff}$ is the ground state energy of $\hat{H}_\mathrm{eff}$ with the correct electron number and spin. We developed a technique to evaluate the logarithm of the Trotter timestep operator, $\exp(\hat{H}_\mathrm{eff}\tau)$, and track its eigenvalues to avoid problems with convergence of the BCH equation (see Appendix~\ref{app:logarithm}). Using a sufficiently small initial $\Delta t$, we apply a constant phase corresponding to $-E_\mathrm{CI}$ and evaluate overlaps of eigenvectors for subsequent values of $\Delta t$ to ensure continuity.

Fig.~\ref{fig:log} shows $\Delta E$ vs. $\Delta t$ for the three operator pools with five random orders of operators using first-order and second-order Trotterization (FOT and SOT, respectively) to give a qualitative understand of the errors. As was also seen for the infidelities, the symmetry conserving Slater and CSF pools show accumulation of error roughly an order of magnitude greater than the Qubit pool with FOT for LiH and H$_2$O. The errors are almost exclusively negative with increasing $\Delta t$, which is explained by the fact that the $\mathcal{O}(\tau)$ error in the BCH expansion yields an antisymmetric matrix multiplied by $i$ in a basis where $\hat{H}$ is diagonal, thus the ground-state energy only decreases in energy. The SOT errors, in contrast, are mostly positive and are comparable in magnitude for all three pools. In general, we found that the $\mathcal{O}(\tau)$ and $\mathcal{O}(\tau^2)$ terms of the BCH expansion were insufficient to estimate the behaviour of the energy error. Fig.~\ref{fig:log}c shows the rapid onset of error for H$_2$O with a STO-3G basis and the Qubit operator pool. The error is highly non-linear in $\Delta t$, and the overall magnitude is much greater than the differences between samples.

For smaller basis sets, H$_2$ and LiH also show error magnitudes that are comparable for FOT and SOT (see Appendix~\ref{app:ordering}). Altogether, the trends in our calculations suggest that larger molecules show a benefit from SOT that outweighs the errors from the symmetry conserving operator pools. Conservation of total spin increases the number of exponentials of Hermitian excitation operators from 4 to 11 for $\hat{C}_q^{pr}$ using operator kirigami, but for larger molecules the increase in the $\mathcal{O}(N_\mathrm{orb}^3)$ terms from $\hat{C}_q^{pr}$ is negligible compared to the $\mathcal{O}(N_\mathrm{orb}^4)$ terms in the Hamiltonian with $p > q > r > s$. Finding and adopting symmetry conserving operator pools for quantum computation thus provides the advantage of re-purposing optimized electronic structure methods to test quantum computing algorithms at a much greater scale than previous tests.

In this work, we showed that fermion-to-qubit encodings of the electronic Hamiltonian share a common set of Hermitian excitation operators which conserve symmetries of the Hamiltonian with Abelian groups of symmetry operators when Trotterization is employed. Symmetries with non-Abelian groups, however, are not conserved by all products of exponentials of Hermitian excitation operators. To achieve non-Abelian symmetry conservation, we developed operator kirigami, which cuts a sum of Hermitian excitation operators into commuting and non-commuting components and folds the non-commuting components into a single operator. We applied operator kirigami to achieve electron number and total spin symmetry conservation. Finally, we investigated the consequences of symmetry conservation for Trotterized time evolution and quantum phase estimation. Errors were roughly an order of magnitude greater for symmetry conserving operator pools with first-order Trotterization, but second-order Trotterization showed comparable errors for the larger systems tested. We have developed and demonstrated a technique capable of reducing removing Trotter errors, including symmetry breaking. Furthermore, our work levels the computational effort of the simulation quantum computing methods on classical computers to that of established classical computing methods, enabling the development, testing, and eventual deployment of quantum computing algorithms on a chemically relevant scale.

\vspace{-4ex}
\section*{Acknowledgements}
We would like to thank Peter Selinger and Julien Ross for the insightful discussion. We thank the Atlantic Computing Excellence Network (ACENET) for computational resources. E. L. S. and R. J. M. were supported by funding from the Defence Advanced Research Projects Agency IMPAQT program (HR-0011-24-3-0301) and by the National Science and Engineering Research Council through an NSERC Alliance grant (ALLRP 592521-23). J.-K. H. and R. J. M. were supported by funding from National Research Council Canada through the Applied Quantum Challenge program (AQC-100).

\vspace{4ex}
\section*{Data Availability}
The data needed to reproduce the findings of this article are provided in the Supplemental Material.

\bibliography{biblio}

\onecolumngrid

\section*{APPENDICES}
\appendix

\section{Explicit form of the electronic Hamiltonian}
\label{app:integham}
The electronic Hamiltonian is typically written with second quantization as
\begin{equation} \label{eq:1ham}
    \hat{H} = \sum_{pq} h_q^p \hat{a}_p^\dag \hat{a}_q + \frac{1}{2}\sum_{pqrs} v_{rs}^{pq} \hat{a}_p^\dag \hat{a}_q^\dag \hat{a}_s \hat{a}_r = \sum_{pq} h_q^p \hat{a}_p^\dag \hat{a}_q - \frac{1}{2}\sum_{pqrs} v_{rs}^{pq} \hat{a}_p^\dag \hat{a}_q^\dag \hat{a}_r \hat{a}_s,
\end{equation}
where (without spin-orbit couplings) the one- and two-electron integral coefficients are given by
\begin{align}
    h_q^p &= \braket{p|\hat{h}|q} = -\int\mathrm{d}\mathbf{x}_1\,\phi_p^*(\mathbf{x}_1)\left(\frac{1}{2}\bm{\nabla}_1^2 + \sum_\alpha \frac{Z_I}{R_{1I}}\right)\phi_q(\mathbf{x}_1) \nonumber\\
    &= -\int\mathrm{d}\mathbf{r}_1\,\phi_p^*(\mathbf{r}_1)\left(\frac{1}{2}\bm{\nabla}_1^2 + \sum_I \frac{Z_I}{R_{1I}}\right)\phi_q(\mathbf{r}_1) \int\mathrm{d}s_1\, \sigma_p^*(s_1) \sigma_q(s_1) = \tilde{h}_q^p \delta_{\sigma_p\sigma_q}, \\
    v_{rs}^{pq} &= \braket{pq|rs} = \int\mathrm{d}\mathbf{x}_1 \mathrm{d}\mathbf{x}_2\,\phi_p^*(\mathbf{x}_1) \phi_q^*(\mathbf{x}_2) \frac{1}{r_{12}} \phi_r(\mathbf{x}_1) \phi_s(\mathbf{x}_2) \nonumber\\
    &= \int\mathrm{d}\mathbf{r}_1 \mathrm{d}\mathbf{r}_2\,\phi_p^*(\mathbf{r}_1) \phi_q^*(\mathbf{r}_2) \frac{1}{r_{12}} \phi_r(\mathbf{r}_1) \phi_s(\mathbf{r}_2) \int\mathrm{d}s_1\, \sigma_p^*(s_1) \sigma_r(s_1) \int\mathrm{d}s_2\, \sigma_q^*(s_2) \sigma_s(s_2) \nonumber\\
    &= \tilde{v}_{rs}^{pq} \delta_{\sigma_p \sigma_r} \delta_{\sigma_q \sigma_s}.
\end{align}
$\mathbf{x}_j$ are electronic coordinates for electron $j$ with a vector of three spatial coordinates $\mathbf{r}_j$ and one spin coordinate $s_j$, $\phi_p(\mathbf{x}_j)$ is the spin orbital $p$, comprised of a product of a spatial orbital $\phi_p(\mathbf{r}_j)$ and a spin function $\sigma_p(s_j)$, $\bm{\nabla}_1^2 = \partial^2/\partial\mathbf{r}_1^2$ is the electronic Laplacian, $Z_I$ is charge of atomic nucleus $I$, $R_{1I} = |\mathbf{r}_1 - \mathbf{R}_I|$ is the distance between electron $i$ and nucleus $I$, and $r_{12} = |\mathbf{r}_1 - \mathbf{r}_2|$ is the inter-electronic distance. We use $\tilde{h}_q^p$ and $\tilde{v}_{rs}^{pq}$ to denote spatial integrals, and $\delta_{\sigma_p \sigma_q}$ for spin function integrals.

Without an external magnetic field, all spin orbitals can be real-valued meaning $h_q^p = h_p^q$, $v_{rs}^{pq} = v_{sr}^{qp}$, and $v_{rs}^{pq} = v_{rq}^{ps} = v_{ps}^{rq} = v_{pq}^{rs}$. We can re-write the Hamiltonian in terms of Hermitian excitation operators $\hat{\Xi}_q^p = \hat{a}_p^\dagger \hat{a}_q + \hat{a}_q^\dagger \hat{a}_p$ by commuting $\hat{a}_q^\dagger$ and $\hat{a}_r$ in Eq.~\ref{eq:1ham}, giving
\begin{equation}
    \hat{H} = \frac{1}{2}\sum_{pq} k_q^p \hat{\Xi}_q^p + \frac{1}{16}\sum_{pqrs} v_{rs}^{pq} \{\hat{\Xi}_r^p, \hat{\Xi}_s^q\},
\end{equation}
where $k_q^p = h_q^p - \sum_r v_{rq}^{pr}$. To take symmetries into account, we assume that the orbitals are symmetry adapted, meaning each orbital corresponds to a single irreducible representation (irrep). As a result, $k_q^p$ vanishes unless the irreps of orbitals $q$ and $p$ match, and $v_{rs}^{pq}$ vanishes unless the direct product of irreps $p$, $q$, $r$ and $s$ includes the completely symmetric irrep. For degenerate irreps, we can divide the orbitals into equivalent sets within the irrep (e.g. $x$ and $y$ or $x$, $y$ and $z$ for spatial symmetry, $\alpha$ and $\beta$ for spin symmetry) using the additional labels $\sigma$ and $\sigma'$. This yields the Hamiltonian
\begin{equation}
    \hat{H} = \frac{1}{2}\sum_{pq} \tilde{k}_q^p \sum_\sigma \hat{\Xi}_{q\sigma}^{p\sigma} + \frac{1}{16} \sum_{pqrs} \bigg[\tilde{u}_{rs}^{pq} \sum_\sigma \{\hat{\Xi}_{r\sigma}^{p\sigma},\hat{\Xi}_{s\sigma}^{q\sigma}\} + \tilde{v}_{rs}^{pq} \sum_{\sigma' \neq \sigma} \{\hat{\Xi}_{r\sigma}^{p\sigma},\hat{\Xi}_{s\sigma'}^{q\sigma'}\} + \tilde{w}_{rs}^{pq} \sum_{\sigma' \neq \sigma} \{\hat{\Xi}_{r\sigma'}^{p\sigma},\hat{\Xi}_{s\sigma}^{q\sigma'}\}\bigg],
\end{equation}
where $\tilde{k}_q^p = k_{q\sigma}^{p\sigma}$, $\tilde{u}_{rs}^{pq} = v_{r\sigma s\sigma}^{p\sigma q\sigma}$, $\tilde{v}_{rs}^{pq} = v_{r\sigma' s\sigma}^{p\sigma' q\sigma}$ and $\tilde{w}_{rs}^{pq} = v_{r\sigma' s\sigma}^{p\sigma q\sigma'}$ are the unique integral coefficients when $\sigma \neq \sigma'$. In general, different orbital indices $p\sigma$ and $q\sigma'$ will have values of $\sigma$ and $\sigma'$ that depend on $p$ and $q$. However, terms with the same $\sigma$ involve orbitals in the same irrep if they have a non-zero coefficient.

In the particular case of electronic spin symmetry, $\tilde{u}_{rs}^{pq} = \tilde{v}_{rs}^{pq}$ and $\tilde{w}_{rs}^{pq} = 0$. Each orbital belongs to the same doubly degenerate irrep which is divided into spins $\alpha$ and $\beta$. The simplified form is given by Eq.~\ref{eq:ham} in the main text. Using the definition $\hat{A}_q^p = \sum_\sigma \hat{\Xi}_{q\sigma}^{p\sigma}$, we can write the Hamiltonian as
\begin{equation}
    \hat{H} = \frac{1}{2} \sum_{pq} \tilde{k}_q^p \hat{A}_q^p + \frac{1}{16} \sum_{pqrs} \tilde{v}_{rs}^{pq} \{\hat{A}_r^p, \hat{A}_s^q\},
\end{equation}
from which we can rewrite the Hamiltonian to reveal all unique terms,
\begin{align} \label{eq:ham_g_fock}
    \hat{H} &= \frac{1}{2}\sum_p \left(\tilde{k}_p^p \hat{A}_p^p + \frac{1}{4}\tilde{v}_{pp}^{pp} \hat{A}_p^p \hat{A}_p^p\right) \nonumber\\
    &\quad+ \sum_{p>q}\left(\tilde{k}_q^p \hat{A}_q^p + \frac{1}{2}\left[\tilde{v}_{qq}^{pp}\hat{A}_q^p \hat{A}_q^p + \tilde{v}_{qq}^{pq}\hat{B}_q^p + \tilde{v}_{qp}^{pp}\hat{B}_p^q\right] + \frac{1}{4}\tilde{v}_{pq}^{pq}\hat{A}_p^p \hat{A}_q^q\right) \nonumber\\
    &\quad+ \sum_{p>q>r}\left(\tilde{v}_{rr}^{pq}\hat{C}_{r}^{pq} + \tilde{v}_{qq}^{pr}\hat{C}_{q}^{pr} + \tilde{v}_{pp}^{qr}\hat{C}_{p}^{qr} + \frac{1}{2}\left[\tilde{v}_{qr}^{pr}\hat{A}_q^p \hat{A}_r^r + \tilde{v}_{rq}^{pq}\hat{A}_r^p \hat{A}_q^q + \tilde{v}_{rp}^{qp}\hat{A}_r^q \hat{A}_p^p\right]\right) \nonumber\\
    &\quad+ \sum_{p>q>r>s} \left(\tilde{v}_{rs}^{pq} \hat{A}_r^p \hat{A}_s^q + \tilde{v}_{sr}^{pq}\hat{A}_s^p \hat{A}_r^q + \tilde{v}_{qs}^{pr}\hat{A}_q^p \hat{A}_s^r \right),
\end{align}
which uses the definitions $\hat{B}_q^p = \frac{1}{2}\{\hat{A}_q^p, \hat{A}_q^q\}$ and $\hat{C}_r^{pq} = \frac{1}{2}\{\hat{A}_r^p, \hat{A}_r^q\}$ for operators made of non-commuting products.

\section{Properties of Hermitian excitation operators} \label{app:prop}
The following properties define the products of Hermitian excitation operators assuming $p \neq q \neq r$, from which their commutators and anticommutators can be derived:
\begin{align}
    (\hat{\Xi}_q^p)^2 &= (\hat{\Sigma}_q^p)^2 = \frac{1}{2}\left(1 - (1 - \hat{\Xi}_p^p)(1 - \hat{\Xi}_q^q)\right) = \frac{1}{2}\left(\hat{\Xi}_p^p + \hat{\Xi}_q^q - \hat{\Xi}_p^p\hat{\Xi}_q^q\right), \\
    (\hat{\Xi}_p^p)^2 &= 2\hat{\Xi}_p^p \quad\Leftrightarrow\quad (1 - \hat{\Xi}_p^p)^2 = 1, \\
    \hat{\Sigma}_q^p \hat{\Xi}_q^p &= -\hat{\Xi}_q^p\hat{\Sigma}_q^p = \frac{i}{2}\left(\hat{\Xi}_p^p - \hat{\Xi}_q^q\right), \\
    \hat{\Xi}_q^p \hat{\Xi}_r^q &= -\hat{\Sigma}_q^p \hat{\Sigma}_r^q = \frac{1}{2}\left(\hat{\Xi}_r^p(1 - \hat{\Xi}_q^q) - i\hat{\Sigma}_r^p\right), \\
    \hat{\Xi}_q^p \hat{\Sigma}_r^q &= \hat{\Sigma}_q^p \hat{\Xi}_r^q = \frac{1}{2}\left(\hat{\Sigma}_r^p(1 - \hat{\Xi}_q^q) + i\hat{\Xi}_r^p\right), \\
    \hat{\Xi}_q^p (1 - \hat{\Xi}_q^q) &= -\hat{\Xi}_q^p (1 - \hat{\Xi}_p^p) = (1 - \hat{\Xi}_p^p)\hat{\Xi}_q^p = -(1 - \hat{\Xi}_q^q)\hat{\Xi}_q^p = i\hat{\Sigma}_q^p.
\end{align}

\section{Operator kirigami} \label{app:kirigami}
In a finite basis of $N_\mathrm{orb}$ orbitals, there are four possible occupations of orbitals $p$ and $q$ when $p \neq q$: those where both are occupied or both are unoccupied, and those where one of the two orbitals is occupied. The Hermitian excitation operators $\hat{\Xi}_q^p = \hat{a}_p^\dagger \hat{a}_q + \hat{a}_q^\dagger \hat{a}_p$ and $\hat{\Sigma}_q^p = i(\hat{a}_p^\dagger \hat{a}_q - \hat{a}_q^\dagger \hat{a}_p)$ are zero when both or neither of the orbitals are occupied. The remaining terms lead to the equation
\begin{equation} \label{eq:1excit}
    \hat{J} = \sum_{j = 1}^{2^{N_\mathrm{orb}-2}} \left(c_j \ket{n_j}\bra{m_j} + c_j^* \ket{m_j}\bra{n_j}\right),
\end{equation}
where $\hat{J} = \hat{\Xi}_q^p$ or $\hat{\Sigma}_q^p$, $c_j = \pm 1$ or $\pm i$, and $\ket{n_j}$ are multi-electron Slater determinants. The matrix has a block-monomial form, meaning $\ket{n_j}$ and $\ket{m_j}$ are unique for each $j$. As a result, the term for each $j$ can be independently diagonalized, yielding eigenvalues of $+1$ and $-1$. $\hat{J}$ is thus tripotent, since its eigenvalues of 0 and $\pm 1$ yield $\hat{J} = \hat{J}^3$.

Based on Eq.~\ref{eq:1excit}, the product of two Hermitian excitation operators is
\begin{equation}
    \hat{J}\hat{K} = \sum_{j,k = 1}^{2^{N_\mathrm{orb}-2}} \left(c_j c_k \delta_{m_j n_k} \ket{n_j} \bra{m_k} + c_j c_k^* \delta_{m_j m_k} \ket{n_j} \bra{n_k} + c_j^* c_k \delta_{n_j n_k} \ket{m_j} \bra{m_k} +  c_j^* c_k^* \delta{n_j m_k} \ket{m_j} \bra{n_k} \right),
\end{equation}
where $\delta$ is the Kronecker delta. Only one of the four terms for each $j,k$ pair can be non-zero, and the same term cannot be non-zero for any $j,k'$ or $j',k$ pair where $j \neq j'$ and $k \neq k'$. The anticommutator $\{\hat{J},\hat{K}\}$ is thus also tripotent, but it has $2^{N_\mathrm{orb}-3}$ terms. Likewise, the commutator multiplied by the imaginary unit, $i[\hat{J},\hat{K}]$, is tripotent with the same number of terms as $\{\hat{J},\hat{K}\}$. By extension, tripotent $n$-electron operators can be constructed using commutators and anticommutators to yield
\begin{equation} \label{eq:nexcit}
    \hat{L} = \sum_{j = 1}^{2^{N_\mathrm{orb}-n-1}} \left(c_j \ket{n_j}\bra{m_j} + c_j^* \ket{m_j}\bra{n_j}\right),
\end{equation}
Operators of this form have the property $\hat{J}\hat{K}\hat{J} = \hat{K}\hat{J}\hat{K} = 0$ when $\hat{J} \neq \hat{K}$ for any $n > 0$.
In a basis where $\hat{\Xi}_q^p$ and $\hat{\Sigma}_q^p$ are off-diagonal for all $p \neq q$, the term $1 - \hat{\Xi}_p^p$ is diagonal and involutory. Multiplication by $1 - \hat{\Xi}_p^p$ thus only changes the sign of elements $c_j$ and $c_j^*$.

The exponential of a tripotent operator is given by
\begin{equation}
    \exp(i\theta\hat{J}) = 1 - \hat{J}^2 + \cos(\theta)\hat{J}^2 + i\sin(\theta)\hat{J}.
\end{equation}
The operator $\hat{J}^2$ is idempotent ($\hat{J}^2 = \hat{J}^{2k}$ for integer $k \neq 0$) and acts as a projector. The exponential thus acts in the space given by $\hat{J}^2$, while the $1 - \hat{J}^2$ space remains unchanged. For a rotation of one operator by another,
\begin{equation}
    \exp(i\theta\hat{J}) \hat{K} \exp(-i\theta\hat{J}) = (1-\hat{J}^2)\hat{K}(1-\hat{J}^2) + \cos(\theta)\{\hat{J}^2,\hat{K}\} + i\sin(\theta)[\hat{J},\hat{K}],
\end{equation}
since $\hat{J}\hat{K}\hat{J} = 0$. In the case where $\hat{L} = i[\hat{J},\hat{K}]$, $1 - \hat{L}^2$ defines the space in which $\hat{J}$ and $\hat{K}$ commute, since $(1 - \hat{L}^2) \hat{L} = \hat{L} - \hat{L}^3 = 0$. Using the exponential to rotate the operators $\hat{J}$ and $\hat{K}$ yields
\begin{align}
    \exp(i\theta\hat{L}) \hat{J} \exp(-i\theta\hat{L}) &= (1-\hat{L}^2)\hat{J}(1-\hat{L}^2) + \cos(\theta)\{\hat{L}^2,\hat{J}\} + \sin(\theta)\{\hat{L}^2,\hat{K}\}, \\
    \exp(i\theta\hat{L}) \hat{K} \exp(-i\theta\hat{L}) &= (1-\hat{L}^2)\hat{K}(1-\hat{L}^2) + \cos(\theta)\{\hat{L}^2,\hat{K}\} - \sin(\theta)\{\hat{L}^2,\hat{J}\},
\end{align}
based on the fact that $[\hat{L},\hat{J}] = -i\{\hat{L}^2,\hat{K}\}$ and $[\hat{L},\hat{K}] = i\{\hat{L}^2,\hat{J}\}$, which are derived by tracking non-zero indices for operators in the form of Eq.~\ref{eq:nexcit}. For any linear combination of $\hat{J}$ and $\hat{K}$ given by $c_J \hat{J} + c_K \hat{K}$ with real coefficients $c_J$ and $c_K$, the non-commuting component of the sum is removed by either of the following
\begin{align}
    \exp(i\theta\hat{L}) (c_J \hat{J} + c_K \hat{K}) \exp(-i\theta\hat{L}) &= (1-\hat{L}^2)(c_J\hat{J} + c_K\hat{K})(1-\hat{L}^2) + \sqrt{c_J^2 + c_K^2} \{\hat{L}^2,\hat{J}\}, \\
    \exp(i\theta'\hat{L}) (c_J \hat{J} + c_K \hat{K}) \exp(-i\theta'\hat{L}) &= (1-\hat{L}^2)(c_J\hat{J} + c_K\hat{K})(1-\hat{L}^2) + \sqrt{c_J^2 + c_K^2} \{\hat{L}^2,\hat{K}\},
\end{align}
where $\theta = \arctan(-c_K/c_J)$ and $\theta' = \arctan(c_J/c_K)$.

We denote this technique ``operator kirigami'' because it has two components: \emph{cutting} the sum by projection into commuting and non-commuting components, and \emph{folding} the non-commuting components into a single operator through unitary rotations. Implementation then involves implementing the commuting component and \emph{unfolding} to generate the non-commuting component.

The specific form of rotation operator ($\hat{L}$) and projector ($\hat{L}^2$) can be simplified because the operators $\hat{J}$ and $\hat{K}$ are invariant with respect to transformations of their null spaces. In the case of $\hat{G}_+ = \hat{\Xi}_{q\alpha}^{p\alpha}\hat{\Xi}_{r\beta}^{q\beta}$, $\hat{G}_- = \hat{\Xi}_{r\alpha}^{q\alpha}\hat{\Xi}_{q\beta}^{p\beta}$ and $\hat{G}_{0\sigma} = \hat{\Xi}_{r\sigma}^{p\sigma}(1 - \hat{\Xi}_{q\sigma}^{q\sigma})$, we first choose
\begin{equation}
    \hat{L}_+ = i[\hat{G}_+,\hat{G}_-] = i[\hat{\Xi}_{q\alpha}^{p\alpha},\hat{\Xi}_{r\alpha}^{q\alpha}] \hat{\Xi}_{r\beta}^{q\beta}\hat{\Xi}_{q\beta}^{p\beta} + i\hat{\Xi}_{q\alpha}^{p\alpha} \hat{\Xi}_{r\alpha}^{q\alpha}[\hat{\Xi}_{r\beta}^{q\beta},\hat{\Xi}_{q\beta}^{p\beta}] = \frac{1}{2}\left(\hat{\Sigma}_{r\alpha}^{p\alpha} \hat{\Xi}_{r\beta}^{p\beta}(1 - \hat{\Xi}_{q\beta}^{q\beta}) - \hat{\Xi}_{r\alpha}^{p\alpha}(1 - \hat{\Xi}_{q\alpha}^{q\alpha}) \hat{\Sigma}_{r\beta}^{p\beta}\right),
\end{equation}
which yields $\hat{L}_+^2 = \frac{1}{8}\prod_\sigma (\hat{\Xi}_{p\sigma}^{p\sigma} + \hat{\Xi}_{r\sigma}^{r\sigma} - \hat{\Xi}_{p\sigma}^{p\sigma}\hat{\Xi}_{r\sigma}^{r\sigma}) - \frac{1}{8}\prod_\sigma (\hat{\Xi}_{p\sigma}^{p\sigma} - \hat{\Xi}_{r\sigma}^{r\sigma})(1 - \hat{\Xi}_{q\sigma}^{q\sigma})$. Performing operator kirigami on $\hat{G}_+ + \hat{G}_-$ yields $(1-\hat{L}_+^2)(\hat{G}_+ + \hat{G}_-)(1-\hat{L}_+^2) + \sqrt{2}\{\hat{L}_+^2,\hat{G}_+\}$, but the same operation on $\hat{G}_0 = (\hat{G}_{0\alpha} + \hat{G}_{0\beta}) / 2$ yields
\begin{align}
    \exp\left(-i\frac{\pi}{4}\hat{L}_+\right) &\hat{G}_0 \exp\left(i\frac{\pi}{4}\hat{L}_+\right) = (1 - \hat{L}_+^2)\hat{G}_0(1 - \hat{L}_+^2) + \frac{1}{\sqrt{2}}\left(\{\hat{L}_+^2,\hat{G}_0\} - i[\hat{L}_+,\hat{G}_0]\right) \nonumber\\
    &= (1 - \hat{L}_+^2)\hat{G}_0(1 - \hat{L}_+^2) + \frac{1}{\sqrt{2}}\left(\hat{\Gamma}_\alpha + \hat{\Gamma}_\beta\right), \\
    \hat{\Gamma}_\alpha &= \frac{1}{4}\hat{G}_{0\alpha} \left(\hat{\Xi}_{r\beta}^{r\beta}(2 - \hat{\Xi}_{p\beta}^{p\beta} - \hat{\Xi}_{q\beta}^{q\beta}) + \hat{\Xi}_{p\beta}^{p\beta}\hat{\Xi}_{q\beta}^{q\beta}\right), \\
    \hat{\Gamma}_\beta &= \frac{1}{4}\hat{G}_{0\beta} \left(\hat{\Xi}_{p\alpha}^{p\alpha}(2 - \hat{\Xi}_{q\alpha}^{q\alpha} - \hat{\Xi}_{r\alpha}^{r\alpha}) + \hat{\Xi}_{q\alpha}^{q\alpha}\hat{\Xi}_{r\alpha}^{r\alpha}\right).
\end{align}
Notably, $\hat{G}_0$ is not tripotent, whereas $\hat{\Gamma}_\alpha + \hat{\Gamma}_\beta$ is tripotent but not in the form of Eq.~\ref{eq:nexcit}, so their commutators with $\hat{G}_+$ cannot be removed by the strategy described above. We could proceed by sequentially using operator kirigami to simplify $\sqrt{2}\{\hat{L}_+^2,\hat{G}_+\} + \hat{\Gamma}_\alpha$, followed by incorporating $\hat{\Gamma}_\beta$. Instead, we found that we can use a simplified folding operator $\hat{R}_+ = \hat{\Sigma}_{r\alpha}^{p\alpha} \hat{\Xi}_{r\beta}^{p\beta}(1 - \hat{\Xi}_{q\beta}^{q\beta})$, which yields
\begin{equation}
    \exp\left(-i\frac{\pi}{4}\hat{R}_+\right) \hat{G}_0 \exp\left(i\frac{\pi}{4}\hat{R}_+\right) = (1 - \hat{R}_+^2)\hat{G}_0(1 - \hat{R}_+^2) + \hat{\Gamma}_\beta.
\end{equation}
The change of operator does not change the rotation of $\hat{G}_+ + \hat{G}_-$, and $\hat{R}_+^2 = \frac{1}{4}\prod_\sigma (\hat{\Xi}_{p\sigma}^{p\sigma} + \hat{\Xi}_{r\sigma}^{r\sigma} - \hat{\Xi}_{p\sigma}^{p\sigma}\hat{\Xi}_{r\sigma}^{r\sigma})$ includes $\hat{L}_+^2$. The $\hat{\Gamma}_\beta$ operator has the form of an excitation operator multiplied by a projection, and is thus tripotent. Following the same procedure, we find
\begin{equation}
    \hat{L}_0 = i[\{\hat{R}_+^2,\hat{G}_+\},\hat{\Gamma}_\beta] = i[\hat{G}_+,\hat{\Gamma}_\beta] = -\frac{1}{2}\left(\hat{\Sigma}_{q\alpha}^{p\alpha} (1 - \hat{\Xi}_{r\alpha}^{r\alpha}) \hat{\Xi}_{q\beta}^{p\beta} + \hat{\Xi}_{q\alpha}^{p\alpha} \hat{\Sigma}_{q\beta}^{p\beta}(1 - \hat{\Xi}_{r\beta}^{r\beta})\right).
\end{equation}
As before, we can simplify the fold to a single operator $\hat{R}_0 = \hat{\Sigma}_{q\alpha}^{p\alpha} (1 - \hat{\Xi}_{r\alpha}^{r\alpha})\hat{\Xi}_{q\beta}^{p\beta}$. Operator kirigami using either $\hat{L}_0$ or $\hat{R}_0$ yields $\sqrt{3}\hat{\Gamma}_\beta$ with a magic angle $\theta_m = \arctan(\sqrt{2})$ (with a sign difference to account for the difference in sign of $\hat{L}_0$ and $\hat{R}_0$).

The two projections $\hat{R}_+^2$ and $\hat{R}_0^2$ can be simplified into one by noting that both act in the space given by $\hat{T} = (1 + \prod_\sigma (1 - \hat{\Xi}_{p\sigma}^{p\sigma})(1 - \hat{\Xi}_{r\sigma}^{r\sigma}))/2$. Furthermore, $\hat{T}$ commutes with $\hat{G}_0$, $\hat{G}_+$, and $\hat{G}_-$, meaning $\{\hat{G}_0,\hat{R}_+^2\} = \hat{G}_0\hat{T}$ and $\{\hat{G}_\pm,\hat{R}_+^2\} = \hat{G}_\pm\hat{T}$. Using $\hat{T}$, $\hat{R}_+$ and $\hat{R}_0$, we have thus shown that $\hat{C}_q^{pr} = \hat{G}_0 + \hat{G}_+ + \hat{G}_-$ is given by
\begin{equation}
    \hat{C}_q^{pr} = \hat{C}_q^{pr}(1 - \hat{T}) + \exp\left(i\frac{\pi}{4}\hat{R}_+\right) \exp\left(-i\theta_m\hat{R}_0\right) (\sqrt{3}\hat{\Gamma}_\beta) \exp\left(i\theta_m\hat{R}_0\right) \exp\left(-i\frac{\pi}{4}\hat{R}_+\right),
\end{equation}
where every operation in the rightmost term involves a single excitation operator which maps to a set of commuting Pauli strings. The commutators between $\hat{G}_0(1 - \hat{T})$, $\hat{G}_+(1 - \hat{T})$, and $\hat{G}_-(1 - \hat{T})$ are each zero because $[\hat{G}_+,\hat{G}_-]\hat{T} = [\hat{G}_+,\hat{G}_-]$ and $[\hat{G}_0,\hat{G}_\pm]\hat{T} = [\hat{G}_0,\hat{G}_\pm]$. Because $\hat{T}$ commutes with each fragment, their errors cancel for every order in the Baker-Campbell-Hausdorff equation. Finally, because $\hat{T}$ and $1 - \hat{T}$ commute, the exponential $\exp(-i\theta\hat{C}_q^{pr})$ can be implemented without Trotter error.

\section{Additional data for Trotterized time evolution and QPE energies} \label{app:ap3}
\begin{figure}[p!]
    \centering
    \includegraphics{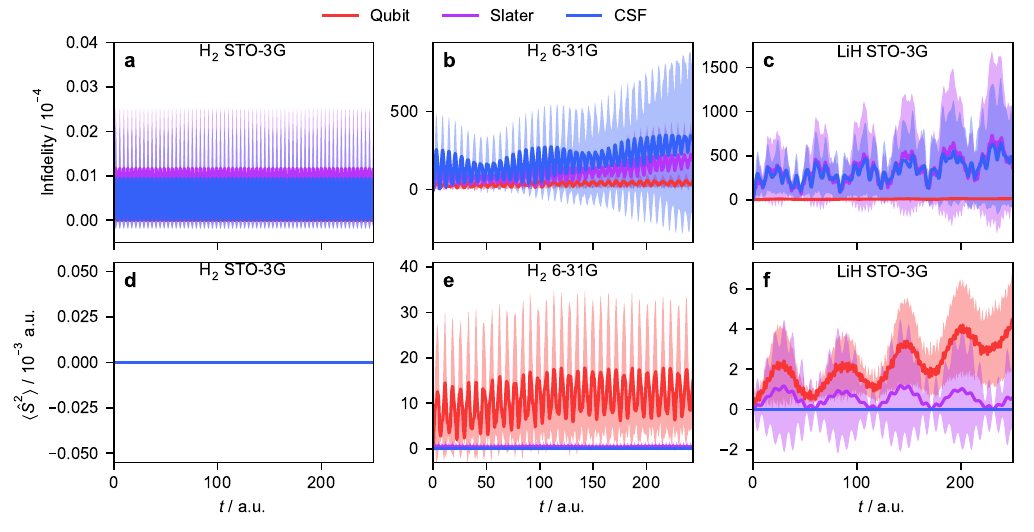}
    \caption{Time evolution of Trotter errors for molecular simulations with the different operator pools. \textbf{a}--\textbf{c}, The average infidelities, $1 - |\braket{\Psi_\mathrm{Trot}|\Psi_\mathrm{exact}}|$, and their standard deviations for 10 random orders of operators. \textbf{d}--\textbf{f}, The average spin squared expectation values, $\braket{\hat{S}^2}$, and their standard deviations for 10 random orders of operators. Columns represent different molecules and basis sets: H$_2$ with a STO-3G basis (\textbf{a}, \textbf{d}), H$_2$ with a 6-31G basis (\textbf{b}, \textbf{e}), and LiH with a STO-3G basis (\textbf{c}, \textbf{f}). }
    \label{fig:ap_fid}
\end{figure}

\begin{figure}[p!]
    \centering
    \includegraphics{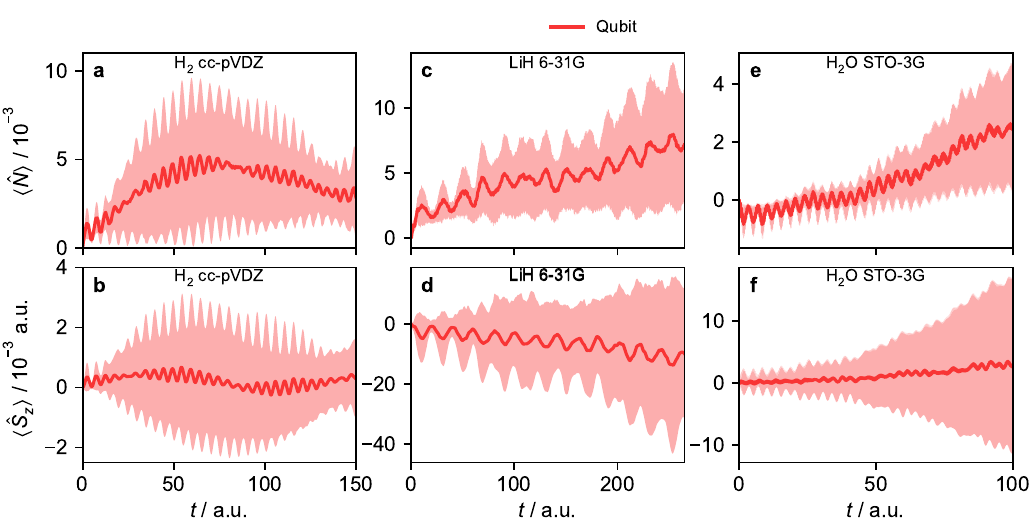}
    \caption{Time evolution of $\hat{N}$ and $\hat{S}_z$ for molecular simulations with the Qubit operator pool. \textbf{a}--\textbf{c}, The average electron number, $\langle \hat{N} \rangle$, and its standard deviation for 10 random orders of operators. \textbf{d}--\textbf{f}, The average spin $z$-projection expectation values, $\braket{\hat{S}_z}$, and their standard deviations for 10 random orders of operators. Columns represent different molecules and basis sets: H$_2$ with a cc-pVDZ basis (\textbf{a}, \textbf{d}), LiH with a 6-31G basis (\textbf{b}, \textbf{e}), and H$_2$O with a STO-3G basis (\textbf{c}, \textbf{f}). The magnitude of this properties is a measure of how many non-physical states populate with time in quantum simulations with the Qubit operator pool.
    }
    \label{fig:n_sz_1}
\end{figure}

\begin{figure}[h!]
    \centering
    \includegraphics{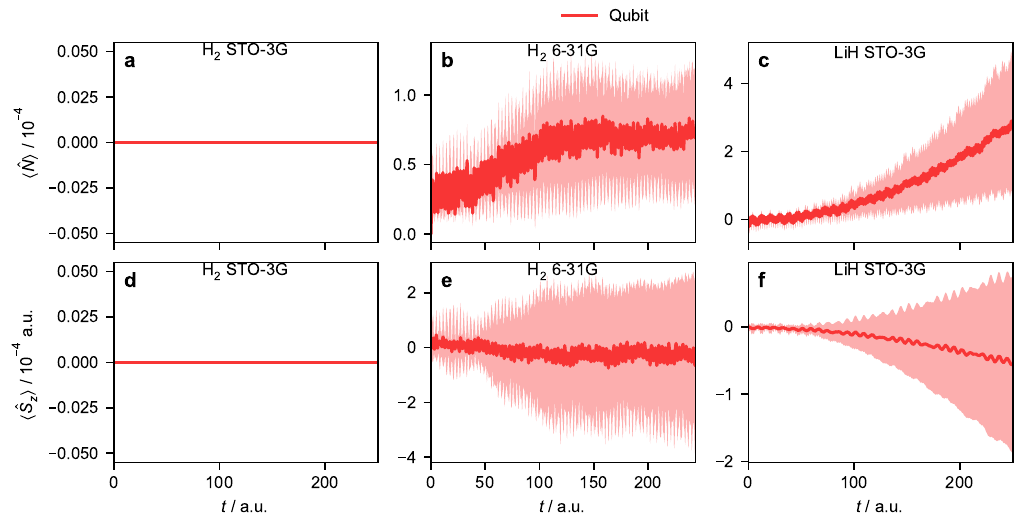}
    \caption{Time evolution of $\hat{N}$ and $\hat{S}_z$ for molecular simulations with the Qubit operator pool. \textbf{a}--\textbf{c}, The average electron number, $\langle \hat{N} \rangle$, and its standard deviation for 10 random orders of operators. \textbf{d}--\textbf{f}, The average spin $z$-projection expectation values, $\braket{\hat{S}_z}$, and their standard deviations for 10 random orders of operators. Columns represent different molecules and basis sets: H$_2$ with a STO-3G basis (\textbf{a}, \textbf{d}), H$_2$ with a 6-31G basis (\textbf{b}, \textbf{e}), and LiH with a STO-3G basis (\textbf{c}, \textbf{f}). The magnitude of this properties is a measure of how many non-physical states populate with time in quantum simulations with the Qubit operator pool.
   }
    \label{fig:n_sz_2}
\end{figure}

\begin{figure}[h!]
    \centering
    \includegraphics{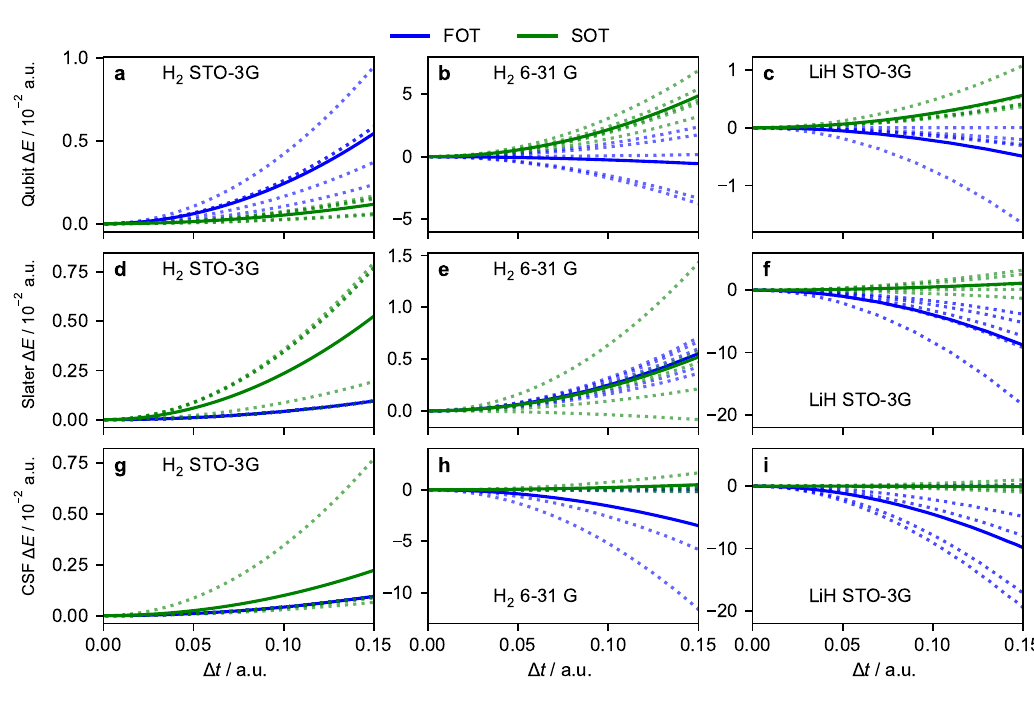}
    \caption{Energy differences between the full CI electronic ground state energy and the Trotter energy for different molecules and operator pools. Five random orders of operators are show with dashed lines, and their average is show as a solid line for first-order Trotter (FOT) and second-order Trotter (SOT). Rows represent different operator pools: Qubit (\textbf{a}--\textbf{c}), Slater (\textbf{d}--\textbf{f}), and CSF (\textbf{g}--\textbf{i}). Columns represent different molecules and basis sets: H$_2$ with a STO-3g basis (\textbf{a}, \textbf{d}, \textbf{g}), H$_2$ with a 6-31G basis (\textbf{b}, \textbf{e}, \textbf{h}), and LiH with a STO-3G basis (\textbf{c}, \textbf{f}, \textbf{i}). All axes share the same units, but note the differences in energy scales.}
    \label{fig:log2}
\end{figure}

Infidelities and $\hat{S}^2$ expectation values are provided in Fig.~\ref{fig:ap_fid} for H$_2$ with STO-3G and 6-31G basis sets, and LiH with a STO-3G basis set. $\hat{N}$ and $\hat{S}_z$ expectation values are shown for H$_2$ with cc-pVDZ, LiH with 6-31G, and H$_2$O with STO-3G in Fig.~\ref{fig:n_sz_1}, and for H$_2$ with STO-3G and 6-31G and LiH with STO-3G in Fig.~\ref{fig:n_sz_2}. Fig.~\ref{fig:log2} shows the QPE energy errors $\Delta E$ vs. $\Delta t$ for H$_2$ with STO-3G and 6-31G, and LiH with STO-3G.

\section{Diagonalization of the logarithm of the Trotter product} \label{app:logarithm}

The differences between the eigenvalues of the effective Hamiltonian, $\hat{H}_\text{eff}$, and the exact Hamiltonian, $\hat{H}$, (see Eq.~\ref{eq:trotter} and \ref{eq:bch}), provide a measure of the error introduced by approximating the time evolution operator with the Trotter formula and its behavior as a function of $\Delta t = i\tau$. Since it is not feasible to expand the BCH formula to infinite order, we recovered the eigenvalues of $\hat{H}_\text{eff}$ by evaluating the logarithm of the product $\prod_j \exp(\tau\hat{H}_j)$. Even though a function for evaluating matrix logarithms is available in the module \texttt{linalg} of the Python library SciPy, applying it directly on the Trotter product can return eigenvalues with an incorrect phase, because of the following relations,
\begin{eqnarray}
    \label{log_angle1}
    \log\left( \exp(\hat{H}_\text{eff}\tau) \right) &=&  \log\left( U \exp(-i\theta_k(\Delta t)) U^\dagger \right) = U\log\left(  \exp(-i\theta_k(\Delta t)) \right) U^\dagger, \\
    \label{log_angle2}
    \log(\exp(-i\theta_k(\Delta t)) ) &=&  \log(\exp(-i\theta_k(\Delta t) + 2\pi n_k ) ).
\end{eqnarray}
Since the complex logarithm is multivalued, SciPy selects by default the principal branch in the complex plane, $\theta_k \in [-\pi,\pi]$, which can cause the value of $\theta_k(\Delta t)$ not to match the value $\Delta t E_k$, with $E_k$ being the energy of the $k$-th eigenstate of $\hat{H}_\text{eff}$. This causes issues especially when the value of $\Delta t E_k$ is much greater than $2\pi$ or it is close to $\pi$. In the first case $E_k\Delta t$ does not belong the principal branch, therefore $\theta_k(\Delta t)$ alone will not recover the $k$-th eigenenergy. For the second case, suppose $\theta_k(\Delta t) \approx \pi$ and it increases with increasing $\Delta t$. There will be a $\Delta t'$ for which $\theta_k(\Delta t') > \pi$, but the direct evaluation of the logarithm will yield $\theta_k(\Delta t') - 2\pi$, since it always returns a value inside the principal branch. This causes discontinuities in $\theta_k$ as a function of $\Delta t$. It is worth noting that, since $\theta_k(\Delta t) / 2\pi $ is not an integer, both situations can occur if the modulus of the division is close to $\pi$.

Here, we used a procedure called branch rotation to recover the eigenvalues of $\hat{H}_\text{eff}$. It consists in multiplying the eigenvalues $\exp(-i\theta_k(\Delta t))$ by an appropriate phase to rotate the angle in which the discontinuity for $\theta_k$ will occur away from $\pi$. Effectively, this selects the branch for $\theta_k$ to be between $[\phi-\pi,\phi+\pi]$. Here, describe how we applied this method to our algorithm for computing the $k$-th eigenvalue of $\hat{H}_\text{eff}$ as a function of $\Delta t$. It is worth mentioning that different eigenvalues can follow different branches, which is equivalent to having different values of $n_k$, therefore the $\phi$ values have to be selected depending on the eigenvalue that is sought.
\begin{enumerate}
    \item Evaluate the Trotter product at a given $-i\Delta t$.
    \item Diagonalize $\exp(-i\hat{H}_\text{eff}\Delta t)$ to obtain $ U \exp(-i\theta_k(\Delta t)) U^\dagger$.
    \item Select the eigenstate $\ket{\phi_{k,\Delta t}}$ as the one with the highest overlap with the k-th eigenvalue of $\exp(-i\hat{H}_\text{eff}\Delta t')$, $\ket{\phi_{k,\Delta t'}}$, where $\Delta t' < \Delta t$. When $\Delta t \rightarrow 0$ the eigenvalue of the exact Hamiltonian, $\ket{\varphi_k}$ can be used as reference.
    \item Multiply $\exp(-i\theta_k(\Delta t))\exp(i\Delta t E_k(\Delta t'))$ where $E_k(\Delta t')$ is the eigenvalue corresponding to $\ket{\phi_{k,\Delta t'}}$. 
    \item Calculate $E_k(\Delta t) =  E_k(\Delta t') - \frac{1}{\Delta t} \text{Arg}( e^{-i\theta_k(\Delta t)}e^{i\Delta t E_k(\Delta t')}  ) $. In Python it is suggested to use \texttt{numpy.angle} to find the argument in the correct quadrant.
    \item Repeat the procedure until the desired time interval is covered.
\end{enumerate}

It is always necessary to have a reference energy and eigenstate, since the value $\Delta t E_k$ determines the branch of the logarithm. For small time steps, the $k$-th eigenstate of the exact Hamiltonian can be used for the steps 3-5 of the procedure. However, for larger time steps, this value can lead to discontinuities, because eigenstates with different energies can belong to different branches. Therefore, it is suggested to use the $k$-th eigenstate of $\hat{H}_\text{eff}$ recovered from a $\Delta t'$ close in value to $\Delta t$.

\section{Evaluation and order of the Trotter products} \label{app:ordering}
In this paper we compare time evolution of physical properties employing the Trotter approximation to the time evolution operator. This approximation can be decomposed to reduce error. Decompositions of order $m$ are used to suppress the error originating from terms of order up to $m-1$. Here we use first and second order to evaluate the effects of suppressing $\mathcal{O}(\tau)$ BCH error for several small molecules and basis sets. The first order formula is given by Eq.~\ref{eq:trotter}, whereas the second order formula is found by:
\begin{eqnarray}
    \label{eq:SOT}
    \exp(\hat{H}\tau) \approx \prod_{j= 0}^{N} e^{-\frac{\tau}{2}\hat{H}_j} \prod_{j=N}^{0} e^{-\frac{\tau}{2}\hat{H}_j} = S_2(\tau).
\end{eqnarray}
We evaluated the logarithm of the two types of products by finding the argument of the eigenvalues of the associated unitary matrix, following steps 2--6 of the algorithm given in the Appendix~\ref{app:logarithm}. This allowed us to find the ground state eigenvalues of effective Hamiltonians, $\hat{H}_\mathrm{eff}$, defined by Eq.~\ref{eq:bch} for the first-order formula. The Hamiltonians associated with the second-order formula also depend on commutators between operator fragments, but does not contain $\mathcal{O}(\tau)$ commutators, $[\hat{H}_i, \hat{H}_j]$ and the coefficients multiplying the remaining terms are smaller than in Eq.~\ref{eq:bch}.

The effective Hamiltonians depend on the sequence in which the exponentials of the fragments are multiplied. To account for the effect of the order on molecular properties, we performed simulations using several distinct exponential sequences. The fragments were evaluated symbolically using the Python library \texttt{OpenFermion}~\cite{openfermion}. For the calculations employing the Qubit pool, we transformed excitation operators using the Jordan-Wigner mapping. We summed all the corresponding Pauli strings into a single sum of \texttt{QubitOperator()} for each molecule. To evaluate the exponentials required for the first- and second-order Trotter product, we first listed the keys of the dictionary formed by calling the attribute \texttt{QubitOperator().terms}, assigning an index to each key, in the order the library provides by default, and randomly shuffled them.  We employed the routine \texttt{get\_sparse\_operator()}, to transform the symbolic expressions of Pauli strings into sparse matrices, computed by the package \texttt{scipy.sparse}. We evaluated the first- and second order Trotter formula employing a different randomly shuffled sequence of indices.

For the operators within the symmetric subspaces, we built the fragments following the order displayed on Eq.~\ref{eq:ham_g_fock}, using nested \texttt{for} cycles to iterate through each combination of $p>q>r>s$. These operations were expressed using the object \texttt{FermionOperator()} and were not transformed to Pauli strings in the numerical simulations. Instead of adding the fragments to form a single \texttt{Fermion Operator()}, we assigned each product of $\hat{A}_q^p$ operators an index and randomly shuffled them for each computation of the electronic dynamics and logarithm evaluations. We developed a function to transform the fragments from their symbolic expression in terms of \texttt{FermionOperator()} to a matrix within the corresponding symmetry-conserving Hilbert space. The fact that they are always expressed as fermionic operators makes our results applicable for any valid fermion-to-qubit transform. In the case of the three-index operators, for the $\hat{N}$ and $\hat{S}_z$ conserving subspace, the exponentials of $\hat{G}_0$ and $\hat{G}_\pm$ fragments were evaluated separately, whereas for the $\hat{S}^2$ conserving subspace, the $C_r^{pq}$ were evaluated using a single exponential for each possible $p,q,r$ combination. This is correct in the classical computation of the quantum algorithm thanks to Eq.~\ref{eq:decomp}. In a quantum computer, the exponential of the l.h.s of this equation is what has to be implemented to conserve spin. For purposes of reproducibility, all of the orderings were saved and are available in the Supplemental Material.


\end{document}